\documentstyle[12pt,epsfig]{article}
\makeatletter
\def\slash#1{{\mathpalette\c@ncel{#1}}} 
\makeatother

\newcommand\beq{\begin{eqnarray}}
\newcommand\eeq{\end{eqnarray}}

\newcommand\la{\langle}
\newcommand\ra{\rangle}

\def\shat{\widehat{s}}

\def\that{\widehat{t}}
\def\uhat{\widehat{u}}

\def\shat{\hat{s}}
\def\that{\hat{t}}
\def\uhat{\hat{u}}

\begin{document}
\begin{flushright}
\end{flushright}
\vspace*{15mm}
\begin{center}
{\Large \bf 
Three-gluon contribution to the single spin asymmetry\\[7pt]
in Drell-Yan and direct-photon processes}
\vspace{1.5cm}\\
 {\sc Yuji Koike$^1$ and Shinsuke Yoshida$^2$}
\\[0.4cm]
\vspace*{0.1cm}{\it $^1$ Department of Physics, Niigata University,
Ikarashi, Niigata 950-2181, Japan}\\
\vspace*{0.1cm}{\it $^2$ Graduate School of Science and Technology, Niigata University,
Ikarashi, Niigata 950-2181, Japan}
\\[3cm]

{\large \bf Abstract} \end{center}
We derive the single-spin-dependent cross section for the 
Drell-Yan lepton-pair production and the direct-photon production
in the $pp$-collision induced by the twist-3 three-gluon
correlation functions in the transversely polarized nucleon in the leading order with respect to
the QCD coupling constant. 
Combined with the contribution from the twist-3 quark-gluon correlation
functions in the  literature, this completes the twist-3 cross section
for these processes.  
We also present a model calculation of the asymmetry for the direct photon
production at the RHIC energy,
demonstrating the sensitivity of the asymmetry to the form of the three-gluon correlation functions.  
In particular, we show that
the asymmetry in the backward direction of the polarized nucleon
is determined by the small-$x$ behavior of the correlation functions.

\newpage
\section{Introduction}
The single transverse-spin asymmetry (SSA) in the
Drell-Yan lepton-pair production and the direct-photon production
provides us with an ideal tool to investigate 
the multi-parton correlations and the orbital motion in the nucleon. (See \cite{review} for a review.)   
The SSAs in these processes are caused solely by those effects in the initial nucleons, while 
the SSAs in hadron productions
in $pp$ collisions and semi-inclusive deep-inelastic scattering (SIDIS)
receive potential contribution also from the multiparton correlations in the fragmentation process.  
For the production of the (virtual) photons with large transverse momentum, 
the process can be analyzed in the framework of the collinear factorization
and the SSA appears as a leading twist-3 observable as a consequence of 
the quark-gluon correlations\,\cite{ET82}-\cite{MSZ11}
and the multi-gluon correlations\,\cite{Ji92}-\cite{KY113} in the transversely polarized nucleon.  
By now there have been several studies on the effect of the quark-gluon correlations
on SSAs in these processes\,\cite{QS92,JQVY06,KT071,KK11,MSZ11}, and the
corresponding twist-3 cross section formula in the leading-order QCD
has been completed\,\cite{KK11}.

As for the multi-gluon correlations, 
the formalism for calculating the single-spin-dependent cross section
induced by the three-gluon correlation functions has recently been established
in \cite{BKTY10,KTY11,KY11}, 
and its contributions to the $D$-meson production
in SIDIS and the $pp$ collision 
have been derived.\footnote{There had been earlier studies on the 
contribution of the three-gluon correlation functions
to SIDIS\,\cite{KQ08} and $pp\to DX$\,\cite{KQVY08}.  However, the formalism used there differs
from \cite{BKTY10} and the result also differs from those in \cite{BKTY10,KY11}. }
Since the gluon-photon and gluon-gluon fusion processes
are the driving processes for the heavy meson productions,
the corresponding SSAs are particularly useful to probe the three-gluon correlations.  
Experimentally, the measurement of the SSA for $D$- and $J/\psi$-
production is ongoing at RHIC in BNL\,\cite{Liu,Jpsi}, from which we hope to extract
the form of the three-gluon correlation functions.
With the use of the extracted three-gluon correlation functions
for the SSAs in the Drell-Yan and the direct photon production in the $pp$ collision,
one can study their impact on the SSA in these processes.
If it turns out that the SSAs in these processes are sensitive to the form 
of the three-gluon correlations, combined analysis of SSAs in $D$-meson productions and
these processes will give more information on the form of 
both quark-gluon and multigluon correlation functions.

The purpose of this paper is to derive the twist-3 single-spin-dependent cross section
for the Drell-Yan and the direct-photon production processes
induced by the three-gluon correlation functions in the transversely polarized nucleon,
applying the formalism developed in \cite{BKTY10}.
\footnote{This paper presents the full detail of the result presented in the conference proceedings \cite{KY112,KY113}. }  
We shall also present a numerical calculation of the asymmetry $A_N^\gamma$ for the
direct photon production using the models for the three-gluon correlation
functions extracted from the RHIC preliminary data of $p^\uparrow p\to DX$.  
We will see that the three-gluon contribution to $A_N^\gamma$ is negligible in the forward
region of the polarized proton but can be substantial in the backward region,
in particular, the $A_N^\gamma$ in the backward region is 
sensitive to the relative sign of the two three-gluon functions and the small-$x$
behavior of the functions.

The rest of this paper is organized as follows:
In Sec. 2, we summarize the complete set of the three-gluon correlation functions in the 
transversely polarized nulceon used in the present analysis.
In Sec. 3, we derive the twist-3 single-spin-dependent cross section
for the Drel-Yan process.  For the derivation, we apply the master formula
developed in \cite{KTY11}.  In Sec. 4, we present the twist-3 cross section for the
direct-photon process by taking the real-photon limit of the result obtained
in Sec. 3.  We also present a model calculation of the asymmetry for this process
$A_N^\gamma\equiv \Delta\sigma^\gamma/\sigma^{\gamma}$ at the RHIC energy.
Sec. 5 is devoted to a brief summary.

\section{Three-gluon correlation functions in the transversely polarized nucleon}

As clarified in \cite{BJLO01,Braun09,BKTY10},
there are two independent three-gluon correlation functions
in the transversely polarized nucleon, 
$O(x_1,x_2)$ and $N(x_1,x_2)$, 
which are the Lorentz-scalar functions of the longitudinal momentum fractions $x_1$ and $x_2$, 
defined as
\beq
&&\hspace{-0.8cm}O^{\alpha\beta\gamma}(x_1,x_2)
=-g(i)^3\int{d\lambda\over 2\pi}\int{d\mu\over 2\pi}e^{i\lambda x_1}
e^{i\mu(x_2-x_1)}\la pS|d_{bca}F_b^{\beta n}(0)F_c^{\gamma n}(\mu n)F_a^{\alpha n}(\lambda n)
|pS\ra \nonumber\\
&&=2iM_N\left[
O(x_1,x_2)g^{\alpha\beta}\epsilon^{\gamma pnS_\perp}
+O(x_2,x_2-x_1)g^{\beta\gamma}\epsilon^{\alpha pnS_\perp}
+O(x_1,x_1-x_2)g^{\gamma\alpha}\epsilon^{\beta pnS_\perp}\right]
\label{3gluonO},\\
&&\hspace{-0.8cm}N^{\alpha\beta\gamma}(x_1,x_2)
=-g(i)^3\int{d\lambda\over 2\pi}\int{d\mu\over 2\pi}e^{i\lambda x_1}
e^{i\mu(x_2-x_1)}\la pS|if_{bca}F_b^{\beta n}(0)F_c^{\gamma n}(\mu n)F_a^{\alpha n}(\lambda n)
|pS\ra \nonumber\\
&&=2iM_N\left[
N(x_1,x_2)g^{\alpha\beta}\epsilon^{\gamma pnS_\perp}
-N(x_2,x_2-x_1)g^{\beta\gamma}\epsilon^{\alpha pnS_\perp}
-N(x_1,x_1-x_2)g^{\gamma\alpha}\epsilon^{\beta pnS_\perp}\right], 
\label{3gluonN}
\eeq
where $F_a^{\alpha\beta}\equiv\partial^\alpha A^\beta_a
-\partial^\beta A^\alpha_a +gf_{abc}A_b^\alpha A_c^\beta$ is the gluon's
field strength, and we used the notation $F_a^{\alpha n}\equiv F_a^{\alpha \beta}n_{\beta}$
and $\epsilon^{\alpha pnS_\perp}\equiv \epsilon^{\alpha\mu\nu\lambda}p_\mu n_\nu S_{\perp\lambda}$
with the convention $\epsilon_{0123}=1$.  
$d^{bca}$ and $f^{bca}$ are the symmetric
and anti-symmetric structure constants of the color SU(3) group,
and we have suppressed the gauge-link operators which ensure the gauge invariance.
$p$ is the nucleon momentum, and
$S_\perp$ is the transverse spin vector of the
nucleon normalized as $S_\perp^2=-1$.
In the twist-3 accuracy, $p$ can be regarded as lightlike ($p^2=0$), 
and $n$ is another lightlike vector satisfying $p\cdot n=1$.  To be specific, 
we set $p^\mu=(p^+,0,\mathbf{0}_\perp)$, $n^\mu=(0,n^-, \mathbf{0}_\perp)$, and  
$S^\mu_\perp =(0,0, \mathbf{S}_\perp)$.
The nucleon mass $M_N$ is introduced to define 
$O(x_1,x_2)$ and $N(x_1,x_2)$ dimensionless.  The
decomposition (\ref{3gluonO}) and (\ref{3gluonN})
takes into account all the constraints from hermiticity, 
invariance 
under the parity- and time-reversal transformations and the permutation 
symmetry among the participating three gluon-fields.  The functions
$O(x_1,x_2)$ and $N(x_1,x_2)$ are real and have the following symmetry
properties,
\beq
&&O(x_1,x_2)=O(x_2,x_1),\qquad O(x_1,x_2)=O(-x_1,-x_2),\label{symO}\\
&&N(x_1,x_2)=N(x_2,x_1),\qquad N(x_1,x_2)=-N(-x_1,-x_2).\label{symN}  
\eeq

\section{Drell-Yan Process}
\subsection{Twist-2 unpolarized cross section}

Before discussing the single-spin-dependent cross section, we
present here the twist-2 unpolarized cross section for the Drell-Yan process,
$p(p)+p(p')\to \gamma^*(q)+X$,
which constitues the denominator of the asymmetry
$A_N^{DY}\equiv \Delta\sigma^{\rm DY}/\sigma^{\rm DY}$.  
The corresponding partonic hard cross section in the leading order QCD (LO) is
obtained from the Feynman diagrams shown in Fig. 1.  
The phase space factor for the virtual photon $d^4q$ is
given in terms of the rapidity $y={1\over 2}\ln\left(q^+/q^-\right)$, 
the squared-mass of the lepton-pair $Q^2=q^2$ and the transverse momentum $\vec{q}_\perp$
as $d^4q={1\over 2} dQ^2 dy d^2\vec{q}_\perp$.
The unpolarized cross section is obtained by taking the trace of the hadronic tensor
$W^{\mu\nu}(p,p',q)$ as 
\beq
{d\sigma^{\rm DY}\over dQ^2dyd^2\vec{q}_{\perp}}&=&{\alpha_{em}^2\over
3\pi SQ^2}(-W^{\mu}_{\ \mu}(p,p',q)) \nonumber\\
&=&{\alpha_{em}^2\alpha_s\over 3\pi SQ^2}
\int{dx\over x}\int{dx'\over x'}
\sum_q e_q^2
\left[f_q(x)f_{\bar{q}}(x')\hat{\sigma}_{q\bar{q}} +G(x)f_q(x')\hat{\sigma}_{gq}
\right.\nonumber\\
&&
\left.+f_q(x)G(x')\hat{\sigma}_{qg}
\right]
\delta(\hat{s}+\hat{t}+\hat{u}-Q^2),
\label{twist2}
\eeq
where $S$ is the squared center-of-mass energy,
$\alpha_{em}\simeq 1/137$ is the QED coupling constant,
$\alpha_s=g^2/(4\pi)$ is the strong coupling constant,   
$e_u=2/3$ and $e_d=-1/3$ {\it etc} are the electric charge of each quark flavor, and 
$f_q(x)$ and $G(x)$ are, respectively, the unpolarized quark distribution with flavor $q$ 
and the gluon distribution.  $\sum_q$ denotes the sum over all quark and antiquark flavors.  
In (\ref{twist2}), $\hat{s}$, $\hat{t}$ and $\hat{u}$ are the 
Mandelstam
variables in the parton level defined as
\beq
\hat{s}=(xp+x'p')^2,\qquad \hat{t}=(xp-q)^2,\qquad \hat{u}=(x'p'-q)^2, 
\label{mandel}
\eeq
and the LO partonic hard cross sections are given by
\beq
\left\{
\begin{array}{lll}
\hat{\sigma}_{q\bar{q}}
=\displaystyle{2C_F\over N}\left({\hat{u}\over
 \hat{t}}+{\hat{t}\over \hat{u}}+{2Q^2\hat{s}\over \hat{t}\hat{u}}\right), \\[0.5cm]
\hat{\sigma}_{gq}=
\displaystyle
-{1\over
 N}\left({\hat{s}\over \hat{u}}+{\hat{u}\over \hat{s}}+{2Q^2\hat{t}\over
 \hat{s}\hat{u}}\right),\\[0.5cm]
 \hat{\sigma}_{qg}=\hat{\sigma}_{gq}(\hat{t}\leftrightarrow\hat{u}),
\end{array}
\right.
\label{tw2hard}
\eeq
where $N=3$ denotes the number of colors for quarks. 
Here $\hat{\sigma}_{q\bar{q}}$ 
and $\hat{\sigma}_{gq}$ are, respectively, obtained from Fig. 1(a) and (b).
In the next section, we will use the relation between
the twist-3 cross section and the $gq\to \gamma^* q$ hard scattering cross section.  
For this purpose, 
we introduce the partonic hard part ${\cal H}^{ab}_{gq,\alpha\beta}$
for the gluon-quark scattering channel
by the relation
\beq
\left(-{1\over 2}g^{\alpha\beta}_{\perp}\right)
{1\over (N^2-1)}\delta_{ab}{\cal H}^{ab}_{gq,\alpha\beta}(xp,x'p',q)=\hat{\sigma}_{gq}
(\hat{s},\hat{t},\hat{u},Q^2)\delta(\hat{s}+\hat{t}+\hat{u}-Q^2),
\label{twist2gq}
\eeq
where
the factors $-1/2g_\perp^{\alpha\beta}$ with $g_\perp^{\alpha\beta}=g^{\alpha\beta} - p^\alpha n^\beta -p^\beta n^\alpha$
and $1/\left(N^2-1\right)\delta_{ab}$ 
are, respectively, associated with the Lorenz and color
projections for the unpolarized gluon density to obtain the cross section
in this channel.  

\begin{figure}[h]
\begin{center}
  \includegraphics[height=5cm,width=7cm]{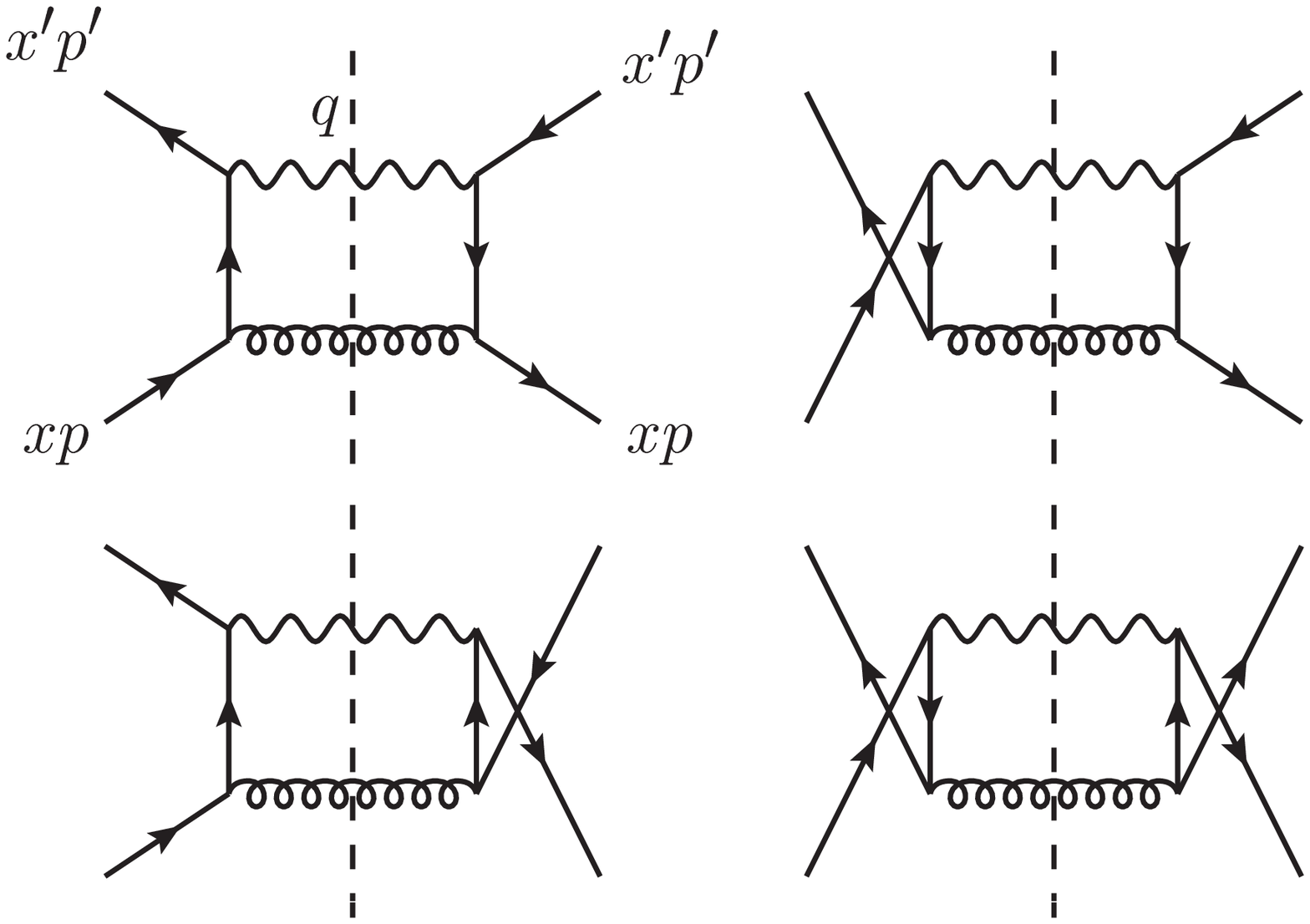}
\hspace{1cm}
  \includegraphics[height=5cm,width=7cm]{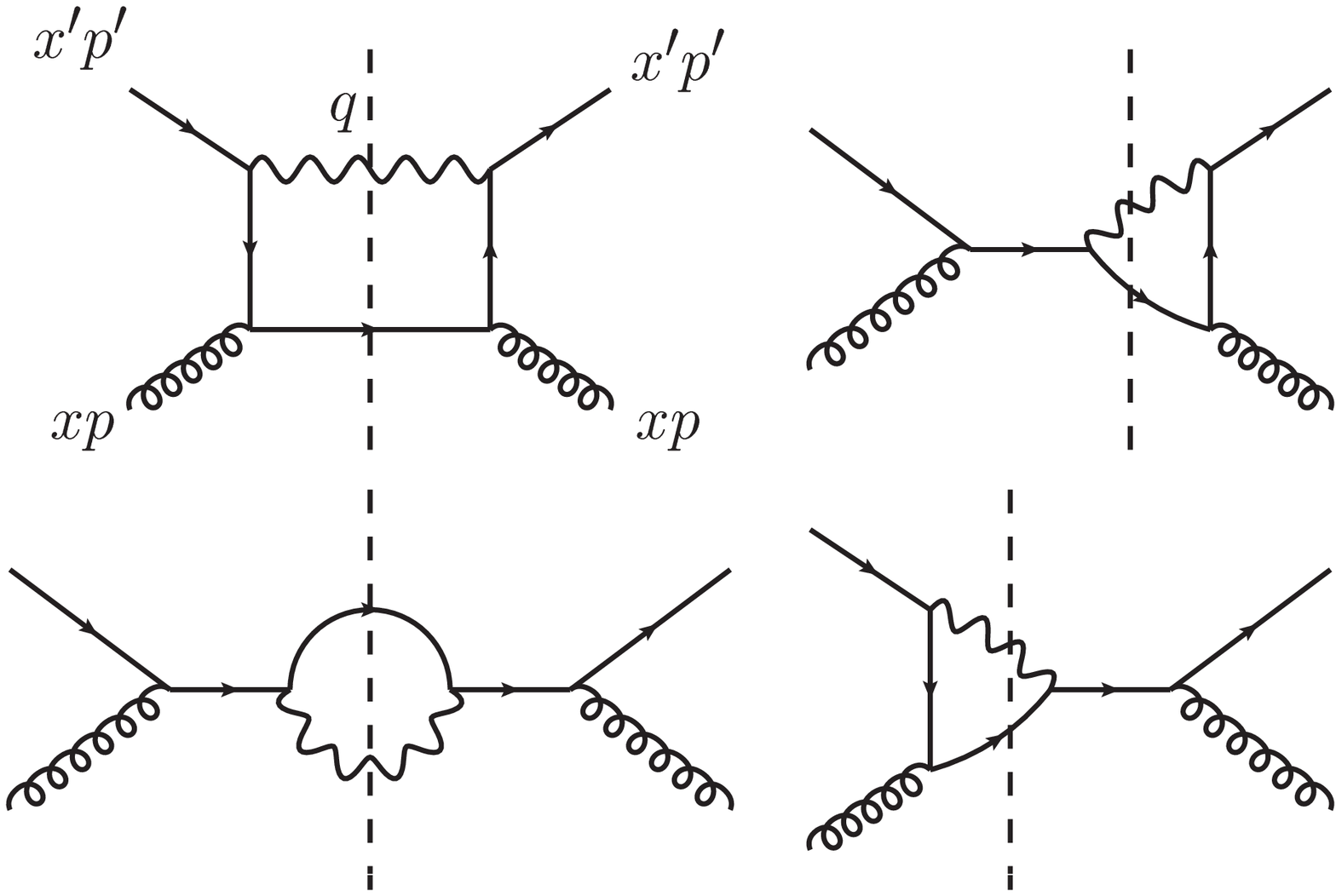}
\end{center}
\hspace{4cm}(a)\hspace{7.5cm}(b)
 \caption{The twist-2 diagrams for the Drell-Yan process.
The diagrams in (a) represent the hard part in the $q\bar{q}$
annihilation channel, and those in (b) represent the hard part
${\cal H}^{ab}_{gq,\alpha\beta}$ in the quark-gluon scattering channel.}
\end{figure}

\subsection{Twist-3 polarized cross section}

The twist-3 cross section for the polarized Drell-Yan process
$p^\uparrow (p,S_\perp)+p(p')\to \gamma^*(q)+X$ is obtained from 
the initial-state-interaction diagrams shown in Fig. 2.  
They represent the hard scattering part $S_{\mu\nu\lambda}^{abc}(k_1,k_2,x'p',q)$
defined as the coefficient for the nucleon matrix element proportional to 
$\la pS_\perp| A_\nu^b A_\lambda^c A_\mu^a|pS_\perp\ra$
where $k_i$ ($i=1,2$) are the momenta coming out of the polarized
nucleon as assigned in Fig. 2.
The spin-dependent cross section is obtained by applying the collinear expansion
to $S_{\mu\nu\lambda}^{abc}(k_1,k_2,x'p',q)$ with respect to $k_i$ around $x_ip$
where $x_i$ denotes the longitudinal momentum fractions.  
The diagrams in Fig. 2 
give rise to the spin-dependent cross section as a pole contribution at $x_1=x_2$  
from the bared propagator which is refered to as the soft-gluon-pole (SGP) contribution.  
Applying the 
formalism developed in \cite{BKTY10}, 
one can obtain the polarized cross section induced by the three-gluon correlation functions
as 
\beq
&&{d\Delta\sigma^{\rm DY}\over dQ^2dyd^2\vec{q}_{\perp}}={\alpha_{em}^2\over
3\pi SQ^2}
(-W^{\mu}_{\ \mu}(p, p',q)) \nonumber\\
&&\qquad\qquad={\alpha_{em}^2\alpha_s\over 3\pi SQ^2}
\sum_q e_q^2 \int\frac{dx'}{x'}f_q(x')
\int\frac{dx_1}{x_1}\int\frac{dx_2}{x_2}
\left[
\left.
{\partial
S_{\mu\nu\lambda}^{abc}(k_1,k_2,x'p',q)p^{\lambda}\over \partial k_2^{\sigma}}\right|_{k_i=x_ip}
\right]^{\rm pole}\nonumber\\
&&\qquad\qquad\qquad\qquad\times\omega^\mu_{\ \,\alpha}\omega^\nu_{\ \,\beta}\omega^\sigma_{\ \,\gamma}
M^{\alpha\beta\gamma}_{F,abc}(x_1,x_2),
\label{DYpol}
\eeq
where $\omega^\mu_{\ \,\alpha}=g^\mu_{\ \,\alpha}-p^\mu n_\alpha$, and 
$M^{\alpha\beta\gamma}_{F,abc}(x_1,x_2)$ is the
lightcone correlation function of the gluon's field-strengths defined as 
\beq
\hspace{-0.5cm}
M^{\alpha\beta\gamma}_{F,abc}(x_1,x_2)
&=&-g(i)^3\int{d\lambda\over 2\pi}\int{d\mu\over 2\pi}e^{i\lambda x_1}
e^{i\mu(x_2-x_1)}\la pS|F_b^{\beta n}(0)F_c^{\gamma n}(\mu n)F_a^{\alpha n}(\lambda n)
|pS\ra \nonumber\\
&=&{N d_{bca}\over (N^2-4)(N^2-1)}
O^{\alpha\beta\gamma}(x_1,x_2)-{if_{bca}\over N(N^2-1)}N^{\alpha\beta\gamma}(x_1,x_2)
\label{Ffunction}
\eeq
with $O^{\alpha\beta\gamma}(x_1,x_2)$ and $N^{\alpha\beta\gamma}(x_1,x_2)$ 
defined in (\ref{3gluonO}) and (\ref{3gluonN}), respectively. 
For convenience, we have factorized the factor $g\alpha_s$ from $S_{\mu\nu\lambda}^{abc}$
and included $g$ and $\alpha_s$ in the correlation function (\ref{Ffunction})
and the prefactor in (\ref{DYpol}), respectively.  
The symbol $[\cdots]^{\rm pole}$ in (\ref{DYpol}) indicates the pole contribution is to be taken from the hard part.  
We emphasize that even though the analysis of Fig. 2 starts with
the gauge-noninvariant correlation function $\la pS_\perp| A_\nu^b A_\lambda^c A_\mu^a|pS_\perp\ra$
and the corresponding hard part $S_{\mu\nu\lambda}^{abc}(k_1,k_2,x'p',q)$, 
gauge-noninvariant contributions appearing in the collinear expansion
either vanish or cancel and 
the total surviving twist-3 contribution to the
single-spin-dependent cross section can be expressed as in (\ref{DYpol}), using the
gauge-invariant correlation functions (\ref{3gluonO}) and (\ref{3gluonN}).

\begin{figure}[ht]
\begin{center}
  \includegraphics[height=5cm,width=8cm]{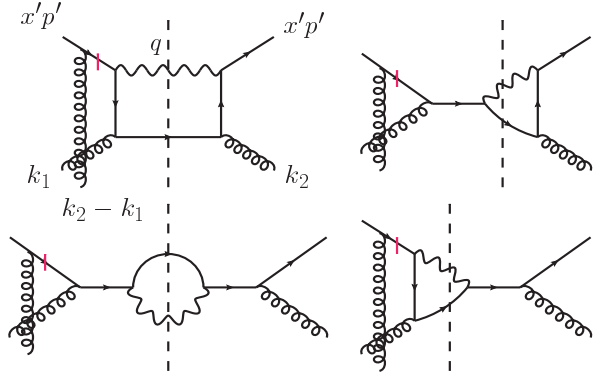}
\end{center}
 \caption{Diagrams representing the three-gluon contribution to the Drell-Yan process.
 The pole part of the bared propagator in each diagram gives rise to the single-spin-dependent cross section.
 The mirror diagrams also contribute. }
\end{figure}

As was shown in \cite{KTY11,KY11}, one can simplifies the actual calculation 
of the twist-3 cross section
by using the ``master formula" developed for the
three-gluon correlation functions.  
This simplification occurs due to the fact that the diagrams in Fig. 2 are obtained 
by attaching the extra gluon-line to those in Fig. 1(b) and
the internal quark-propagator next to this attachment gives the pole contribution.     
For the contribution from the initial-state interaction, 
the derivative
$\left[(\partial/\partial k_2^\sigma)S_{\mu\nu\lambda}^{abc}(k_1,k_2,x'p',q)p^\lambda|_{k_i=x_ip}\right]^{\rm pole}$
is related to the $gq\to \gamma^*q$ hard scattering part shown in Fig. 1 (b) as
\beq
&&\left[\left.{\partial S_{\mu\nu\lambda}^{abc}(k_1,k_2,x'p',q)p^{\lambda}\over 
\partial k_2^{\sigma}}\right|_{k_i=x_ip}\right]^{\rm pole} \nonumber\\
&&\qquad = \left[{-1\over x_2 - x_1 +i\epsilon}\right]^{\rm pole} 
 \left( {\partial \over \partial (x'p'^\sigma)}
-{p'_{\sigma} p^\lambda \over p\cdot p'} {\partial \over \partial (x'p'^\lambda)} \right) 
{\cal H}^{abc}_{gq,\mu\nu}(x_1p,x'p',q)\nonumber\\
&&\qquad = 
i\pi \delta(x_1-x_2)
{d\over d (x'p'^\sigma)}{\cal H}^{abc}_{gq,\mu\nu}(x_1p,x'p',q),
\label{DYmaster}
\eeq
where ${\cal H}^{abc}_{gq,\mu\nu}$ is obtained from ${\cal H}^{ab}_{gq,\mu\nu}$ in (\ref{twist2gq})
by inserting an extra color matrix $t^{c}$ into the place where the extra gluon-line is attached in Fig. 2.  
In the second expression of (\ref{DYmaster}), the partial derivative with respect to $x'p'^\sigma$
implies the shorthand notation of
\beq
{\partial \over \partial (x'p'^\sigma)}f(x'p') = \left.{\partial \over \partial r^\sigma}f(r)\right|_{r\to x'p'}. 
\eeq
In the last expression of (\ref{DYmaster}), the on-shell form of 
$p'^\mu=(p'^+={\vec{p'}_\perp^2\over 2p'^-},p'^-,\vec{p'}_\perp)$ with $\vec{p'}_\perp\neq 0$ should be
used in taking the derivative, regarding $p'^+$ as a dependent variable.  
Since we are considering the cross section in the frame where $p$ and $p'$ are collinear, 
we shall take the limit $\vec{p'}_\perp\to 0$ after performing the derivative $d/dx'{p'}^\sigma$.  
We remind that  ${\cal H}^{abc}_{gq,\mu\nu}(xp,x'p',q)$ 
contains the factor
$\delta((xp+x'p'-q)^2)$ associated with the on-shell condition for the final unobserved parton,
to which the derivative $d/dx'{p'}^\sigma$ also hits.

Substituting (\ref{Ffunction}) and (\ref{DYmaster}) into (\ref{DYpol}), one obtains the cross section as
\beq
{d\Delta\sigma^{\rm DY}\over dQ^2dyd^2\vec{q}_{\perp}}
&=&{\alpha_{em}^2\alpha_s\over 3\pi SQ^2}
\sum_q e_q^2 \int\frac{dx'}{x'}f_q(x')
\int\frac{dx}{x^2}(i\pi) \left.
{d\over d (x'p'^\gamma)}{\cal H}^{abc}_{gq,\alpha\beta}(xp,x'p',q)\right|_{\vec{p'}_\perp\to 0}\nonumber\\
&&\qquad\times\left[{N d_{bca}\over (N^2-4)(N^2-1)}
O^{\alpha\beta\gamma}_\perp(x,x)-{if_{bca}\over N(N^2-1)}N_\perp^{\alpha\beta\gamma}(x,x)\right],
\label{DYpolmas}
\eeq
where
\beq
&&O_\perp^{\alpha\beta\gamma}(x,x)=2iM_N\left[
O(x,x)g_\perp^{\alpha\beta}\epsilon^{\gamma pnS_\perp}
+O(x,0)(g_\perp^{\beta\gamma}\epsilon^{\alpha pnS_\perp}
+g_\perp^{\gamma\alpha}\epsilon^{\beta pnS_\perp})\right],\nonumber\\
&&N_\perp^{\alpha\beta\gamma}(x,x)=2iM_N\left[
N(x,x)g_\perp^{\alpha\beta}\epsilon^{\gamma pnS_\perp}
-N(x,0)(g_\perp^{\beta\gamma}\epsilon^{\alpha pnS_\perp}
+g_\perp^{\gamma\alpha}\epsilon^{\beta pnS_\perp})\right]. 
\label{3gluonxx}
\eeq
From this form, one sees that the twist-3 cross section is written in terms of 
the four functions $O(x,x)$, $O(x,0)$, $N(x,x)$ and $N(x,0)$ as in 
$ep^\uparrow\to eDX$ and $p^\uparrow p\to DX$.

The hard cross sections for $O(x,x)$ and $N(x,x)$ are obtained from the
contraction of ${\cal H}_{gq,\alpha\beta}^{abc}$ with $g_\perp^{\alpha\beta}$ as in the case of the 
unpolarized cross section 
in (\ref{tw2hard}).
Taking the color contraction of ${\cal H}^{abc}_{gq,\alpha\beta}$, one obtains
\footnote{Note the sign of each term in (\ref{3gluonxx}) and (\ref{xx}).}
\beq
&&{Nd_{bca}\over (N^2-1)(N^2-4)}
{\cal H}_{gq,\alpha\beta}^{abc}(xp,x'p',q)g_{\perp}^{\alpha\beta}
={if_{bca}\over N(N^2-1)}
{\cal H}_{gq,\alpha\beta}^{abc}(xp,x'p',q)g_{\perp}^{\alpha\beta}\nonumber\\
&&\qquad= -\hat{\sigma}_{gq}(\hat{s},\hat{t},\hat{u},Q^2)
\delta\left(\hat{s}+\hat{t}+\hat{u}-Q^2\right),
\label{xx}
\eeq
where
$\hat{\sigma}_{gq}$ is the twist-2 unpolarized cross section in the $gq$-channel defined in (\ref{tw2hard}).  
For the scalar functions $\hat{\sigma}_{gq}(\hat{s},\hat{t},\hat{u},Q^2)$ 
and $\delta\left(\hat{s}+\hat{t}+\hat{u}-Q^2\right)$, 
one can perform the derivative with respect to
$x'p'^\gamma$ in (\ref{DYpolmas}) through that with respect to 
$\hat{u}$ as
\beq
&&\left.
{d\over d(x'p'^\gamma)}\hat{\sigma}_{gq}(\hat{s},\hat{t},\hat{u},Q^2)
\delta\left(\hat{s}+\hat{t}+\hat{u}-Q^2\right)\right|_{\vec{p'}_\perp\to 0}\nonumber\\
&&\qquad\qquad=-2q_{\gamma}{\partial \over \partial\hat{u}}
\hat{\sigma}_{gq}(\hat{s},\hat{t},\hat{u},Q^2)
\delta\left(\hat{s}+\hat{t}+\hat{u}-Q^2\right).
\label{scalar}
\eeq
From (\ref{xx}) and (\ref{scalar}), one sees that
the hard cross section for $O(x,x)$ and $N(x,x)$ are common and are
determined completely from the twist-2 partonic cross section in 
the gluon-quark scattering channel.  

The hard cross sections for $O(x,0)$ and $N(x,0)$ are obtained from the
contraction of $\left(d/dx'p'^\gamma\right){\cal H}_{gq,\alpha\beta}^{abc}$ with
$g_\perp^{\beta\gamma}\epsilon^{\alpha pnS_\perp}
+g_\perp^{\gamma\alpha}\epsilon^{\beta pnS_\perp}$.  
To carry out the derivative, 
we introduce the two fixed vectors
$X^\mu =(0,1,0,0)$ and $Y^\mu =(0,0,1,0)$, and write
\beq
g_{\perp}^{\beta\gamma}= -X^\beta X^\gamma -Y^\beta Y^\gamma.  
\eeq
Then the derivative ${d/ dx'p'^\gamma}$ hitting the hard part in (\ref{DYpolmas})
for $O(x,0)$ and $N(x,0)$ can be written as
\beq
&&
{d\over dx'p'^\gamma} {\cal H}_{gq,\alpha\beta}^{abc}(xp,x'p',q)
\left( g_\perp^{\beta\gamma}\epsilon^{\alpha pnS_\perp} + g_\perp^{\alpha\gamma}\epsilon^{\beta pnS_\perp}\right)
\nonumber\\
&&\qquad
=-X^\mu{d\over dx'p'^\mu} {\cal H}_{gq,\alpha\beta}^{abc}(xp,x'p',q)
\left( X^{\beta}\epsilon^{\alpha pnS_\perp} + X^{\alpha}\epsilon^{\beta pnS_\perp}\right)\nonumber\\
&&\qquad\quad -Y^\mu{d\over dx'p'^\mu} {\cal H}_{gq,\alpha\beta}^{abc}(xp,x'p',q)
\left( Y^{\beta}\epsilon^{\alpha pnS_\perp} + Y^{\alpha}\epsilon^{\beta pnS_\perp}\right). 
\label{deriv}
\eeq
To perform the derivative with respect to $x'p'^\gamma$, we note that
the color contraction of ${\cal H}_{gq,\alpha\beta}^{abc}$ can be decomposed by 
introducing the scalar functions $J_i(\hat{s},\hat{t},\hat{u},Q^2)$ ($i=1,\cdots,4$)
as
\beq
&&{Nd_{bca}\over (N^2-1)(N^2-4)}
{\cal H}_{gq,\alpha\beta}^{abc}(xp,x'p',q)
\left(X^\beta \epsilon^{\alpha pnS_\perp} + X^\alpha \epsilon^{\beta
pnS_\perp}\right)\nonumber\\
&&\quad=
{if_{bca}\over N(N^2-1)}
{\cal H}_{gq,\alpha\beta}^{abc}(xp,x'p',q)
\left(X^\beta \epsilon^{\alpha pnS_\perp} + X^\alpha \epsilon^{\beta
pnS_\perp}\right)\nonumber\\
&&\quad= \left[
J_1(\hat{s},\hat{t},\hat{u},Q^2)\left(q\cdot X\right)
\epsilon^{qpnS_\perp}
+J_2(\hat{s},\hat{t},\hat{u},Q^2)  \epsilon^{X
pnS_\perp}
+J_3(\hat{s},\hat{t},\hat{u},Q^2)\left(x'p'\cdot
X\right) \epsilon^{qpnS_\perp}\right.\nonumber\\
&&\left.\qquad\qquad+J_4(\hat{s},\hat{t},\hat{u},Q^2)\left(q\cdot X\right)
x'\epsilon^{p' pnS_\perp}\right]
\delta\left(\hat{s}+\hat{t}+\hat{u}-Q^2\right),
\label{X0}
\eeq
where we have ignored the term proportional to
$(p'\cdot X)\epsilon^{p'pnS_\perp}$, since it vanishes as ${p'}_\perp^\gamma\to 0$
after taking the derivative $d/dx'{p'}^\gamma$.  
The contraction with 
$Y^\beta \epsilon^{\alpha pnS_\perp} + Y^\alpha \epsilon^{\beta
pnS_\perp}$ is similarly decomposed, using the same functions $J_i$ ($i=1,\cdots,4$).  
From (\ref{deriv}) and (\ref{X0}), 
one can perform the derivative
to get the cross section.  
In using the form (\ref{X0}), we remind that 
$J_2$ is dimensionless, while $J_{1,3,4}$ have mass dimension $-2$.  

Using (\ref{xx}), (\ref{scalar}), (\ref{deriv}) and (\ref{X0})
in (\ref{DYpolmas}), and noting that the polarized hard cross section 
for the antiquark contribution changes sign for the $O$
contribution, 
the twist-3 polarized cross section can be expressed as
\beq
&&
{d\Delta\sigma^{\rm DY}\over dQ^2dyd^2\vec{q}_{\perp}}\nonumber\\
&&\quad
={\alpha_{em}^2\alpha_s\over 3\pi SQ^2}(2\pi M_N)
\epsilon^{qpnS_\perp}\sum_q e_q^2\int{dx'\over x'}f_q(x')\int{dx\over
x^2}\,\delta\left( \hat{s}+\hat{t}+\hat{u}-Q^2\right)
\nonumber\\
&&\qquad\times\left[
{2\over\hat{u}}\delta_q
\left\{
Q^2\left({\partial \hat{\sigma}_{gq}\over \partial Q^2}+{\partial \hat{\sigma}_{gq}\over
\partial \hat{t}}\right)O(x,x)
-\hat{\sigma}_{gq}\left( x{dO(x,x)\over dx}-2O(x,x)\right) \right\}\right.\nonumber\\
&&\left.\qquad 
\qquad -\left\{ \delta_q\to 1,\ O(x,x)\to N(x,x)\right\}\right.\nonumber\\
&&\left.\qquad
+{2\over \hat{u}}\delta_q
\left\{ {(\hat{t}-Q^2)(\hat{u}-Q^2) \over \hat{s} }-Q^2\right\}
\left\{
\left( J_1 +Q^2\left({\partial J_1 \over \partial Q^2}+{\partial J_1 \over
\partial \hat{t}}\right)\right)O(x,0)
\right.\right.\nonumber\\
&&\left.\left. \qquad\qquad\qquad
\qquad\qquad\qquad\qquad -J_1\left( x{d O(x,0)\over dx}
-2O(x,0)\right)\right\}
\right.\nonumber\\
&&\left.\qquad
+{2\over \hat{u}}\delta_q
\left\{
-Q^2\left({\partial J_2 \over \partial Q^2}+{\partial J_2 \over \partial
\hat{t}}\right)O(x,0)
+J_2\left( x{d O(x,0)\over dx} -2O(x,0)\right)
\right\}\right.\nonumber\\
&&\left.\qquad
-\delta_q(2J_3+J_4)O(x,0)\right.\nonumber\\
&&\left. \qquad + \left\{ \delta_q \to 1,\  O(x,0)\to
N(x,0)\right\}
\large\right],
\label{masterDY}
\eeq
where $\delta_q=1$ for the quark contribution and $\delta_q=-1$ for the antiquark contribution.
In writing down (\ref{masterDY}),
we have used the relations $\vec{q}_\perp^2= (\hat{t}-Q^2)(\hat{u}-Q^2)/\hat{s}-Q^2$, and 
\beq
&&\hat{s}{\partial J_1\over \partial \hat{s}}+\hat{t}{\partial J_1\over \partial \hat{t}}
+\hat{u}{\partial J_1\over \partial \hat{u}}+ Q^2{\partial J_1\over \partial Q^2}
+J_1=0,\nonumber\\
&&\hat{s}{\partial J_2\over \partial \hat{s}}+\hat{t}{\partial J_2\over \partial \hat{t}}
+\hat{u}{\partial J_2\over \partial \hat{u}}+ Q^2{\partial J_2\over \partial Q^2}=0,
\eeq
which follows from the scale
invariant properties $J_2(\lambda\hat{s},\lambda\hat{t},\lambda\hat{u},\lambda Q^2)
=J_2(\hat{s},\hat{t},\hat{u},Q^2)$ and a similar relation for $Q^2 J_1(\hat{s},\hat{t},\hat{u},Q^2)$.  
By the explicit calculation of $J_i(\hat{s},\hat{t},\hat{u},Q^2)$ ($i=1,\cdots,4$),
one obtaines the twist-3 polarized cross section induced by the three-gluon correlation functions as
\beq
&&{d\Delta\sigma^{\rm DY}\over dQ^2dyd^2\vec{q}_{\perp}}\nonumber\\
&&={2M_N\alpha_{em}^2\alpha_s\over
3 SQ^2}\int {dx\over x}\int {dx'\over x'}\delta
(\hat{s}+\hat{t}+\hat{u}-Q^2)\epsilon^{q p n S_{\perp}}{1\over \hat{u}}\sum_qe_q^2f_q(x') \nonumber\\
&&
\times\left[\delta_q\left\{
\left(\frac{dO(x,x)}{dx}-\frac{2O(x,x)}{x}\right)\hat{\sigma}_1
+\left(\frac{dO(x,0)}{dx}-\frac{2O(x,0)}{x}\right)\hat{\sigma}_2 
+{O(x,x)\over x}\hat{\sigma}_3+{O(x,0)\over x}\hat{\sigma}_4
\right\}\right.\nonumber\\
&&
\left.-\left(\frac{dN(x,x)}{dx}-\frac{2N(x,x)}{x}\right)\hat{\sigma}_1
+\left(\frac{dN(x,0)}{dx}-\frac{2N(x,0)}{x}\right)\hat{\sigma}_2 
-{N(x,x)\over x}\hat{\sigma}_3+{N(x,0)\over x}\hat{\sigma}_4
\right],\nonumber\\
\label{twist3DY}
\eeq
where the partonic hard cross sections are given by
\beq
\left\{
\begin{array}{lll}
\hat{\sigma}_{1}=\displaystyle{2\over N}\left({\hat{u}\over \hat{s}}+{\hat{s}\over
\hat{u}}+{2Q^2\hat{t}\over \hat{s}\hat{u}}\right), \\[0.5cm]
\hat{\sigma}_{2}=\displaystyle{2\over N}\left({\hat{u}\over \hat{s}}+{\hat{s}\over
\hat{u}}+{4Q^2\hat{t}\over \hat{s}\hat{u}}\right), \\[0.5cm]
\hat{\sigma}_{3}=-\displaystyle{1\over N}{4Q^2(Q^2+\hat{t})\over \hat{s}\hat{u}}, \\[0.5cm]
\hat{\sigma}_{4}=-\displaystyle{1\over N}{4Q^2(3Q^2+\hat{t})\over \hat{s}\hat{u}}.
\end{array}
\right.
\label{DYtw3hard}
\eeq
For a large $Q^2$, $\hat{\sigma}_{1}$ differs from $\hat{\sigma}_{2}$ significantly, and
$\hat{\sigma}_{3,4}$ are not negligible.  Therefore the cross section
depends on the four functions $O(x,x)$, $O(x,0)$, $N(x,x)$ and $N(x,0)$ independently
as in the case of $ep^\uparrow \to eDX$\,\cite{BKTY10}.  

Our derivation of (\ref{twist3DY}) and (\ref{DYtw3hard}) is based on the master formula (\ref{DYmaster}).
Alternatively, one can directly calculate the derivative
$\left[(\partial/\partial k_2^\sigma)S_{\mu\nu\lambda}^{abc}(k_1,k_2,x'p',q)p^\lambda|_{k_i=x_ip}\right]^{\rm pole}$
in (\ref{DYpol}) to get the cross section.  We have performed such calculation
and confirmed that the result agrees with (\ref{twist3DY}).

\section{Direct photon production}
\subsection{Spin-dependent cross section}

The cross section formula for the direct photon production can be easily
obtained by taking the $Q^2\to 0$ limit of the result for the 
Drell-Yan process.
The twist-2 unpolarized cross section for $p(p)+p(p')\to\gamma(q)+X$
is given by
\beq 
&&E_{\gamma}\frac{d\sigma^{\gamma}}{d^3\vec{q}}=\frac{\alpha_{em}\alpha_s}{S}
\sum_{q}e_q^2\int\frac{dx'}{x'}\int\frac{dx}{x} 
\left[f_q(x)f_{\bar{q}}(x')\hat{\sigma}_{q\bar{q}}^{\gamma} +G(x)f_q(x')\hat{\sigma}_{gq}^{\gamma}
\right.\nonumber\\
&&
\left.\qquad\qquad\qquad\qquad\qquad\qquad\qquad +f_q(x)G(x')\hat{\sigma}_{qg}^{\gamma}
\right]
\delta \left(\hat{s}+\hat{t}+\hat{u}\right),
\label{DPunpol}
\eeq
where $E_\gamma$ is the energy of the photon and the partonic hard cross sections are defined as
\beq
\hat{\sigma}_{q\bar{q}}^{\gamma}={2C_F\over N}\left({\uhat\over\that}+{\that\over \uhat}\right),\quad
\hat{\sigma}_{gq}^{\gamma}=-{1\over N}\left({\uhat\over \shat}+{\shat\over \uhat}\right), \quad
\hat{\sigma}_{qg}^{\gamma}=-{1\over N}\left({\that\over \shat}+{\shat\over \that}\right),
\label{DPparton}
\eeq
with $\hat{s}$, $\hat{t}$ and $\hat{u}$ in (\ref{mandel}) with $q^2=0$. 
Likewise the twist-3 polarized cross section 
for  $p^\uparrow (p,S_\perp)+p(p')\to\gamma(q)+X$ is given by
\beq
 E_{\gamma}\frac{d^3\Delta\sigma^{\gamma}}{d^3\vec{q}}&=&
\frac{4\alpha_{em}\alpha_s\pi M_N}{S}\sum_{q}e_q^2
\int\frac{dx'}{x'}f_q(x')\int\frac{dx}{x}\delta
(\hat{s}+\hat{t}+\hat{u})\epsilon^{q p n S_{\perp}}\left({-1\over \hat{u}}\right)
\nonumber\\
&&\qquad
\times
\left[\delta_q
\left(\frac{dO(x)}{dx}-\frac{2O(x)}{x}\right) 
-\left( \frac{dN(x)}{dx}-\frac{2N(x)}{x}\right)\right]\hat{\sigma}_{gq}^{\gamma}, 
\label{twist3DP}
\eeq
where
\beq
O(x)=O(x,x)+O(x,0),\qquad N(x)=N(x,x)-N(x,0).
\label{OandN}
\eeq
This result differs from the previous one in \cite{Ji92}, in which the calculation
was not based on the factorization formula (\ref{DYpol}).  
Compared with the Drell-Yan case (\ref{twist3DY}), this cross section has a much simpler form; 
it depends on the three-gluon correlation functions only through 
the combinations of $dO(x)/dx-2O(x)/x$ and $dN(x)/dx-2N(x)/x$, 
accompanying the common partonic hard cross section 
which is the same as the twist-2 hard cross section
for the $gq\to \gamma q$ scattering in (\ref{DPparton}).  
This is in parallel with the twist-3 cross section for $p^\uparrow p\to DX$ in the $m_c\to 0$ limit.  
(At the RHIC energy, the effect of $m_c\neq 0$ is negligible and thus one could regard
the cross section as a function of $O(x)$ and $N(x)$\,\cite{KY11}.)
We also remind that the SGP contribution from the quark-gluon correlation function $G_F(x,x)$ to
$p^\uparrow p\to \pi X$ also appears in the combination $dG_F(x,x)/dx-G_F(x,x)/x$,
whose origin was clearly understood in terms of the master formula\,\cite{KT072}.

From (\ref{twist3DP}), one sees that
if $dO(x)/dx-2O(x)/x$ and $dN(x)/dx-2N(x)/x$ have the same sign and a similar magnitude, the quark contribution
from the unpolarized nucleon cancel between 
the two contributions, and only
the antiquark-contribution is active, which would lead to a small
asymmetry.
On the other hand, if $dO(x)/dx-2O(x)/x$ and $dN(x)/dx-2N(x)/x$ have the opposite sign, 
the quark-contribution become active and thus one would expect a large asymmetry.
Therefore, $A_N^\gamma$ can be an important
measure for the the relative sign and magnitude of the three-gluon correlation functions.

\subsection{Numerical calculation of the asymmetry}

In the previous section, we saw that
the spin-dependent cross section for the direct-photon production
depends on the two combinations $O(x)$ and $N(x)$ in (\ref{OandN}).  
These are the same functions which govern the
SSA for the $p^\uparrow p\to DX$\,\cite{KY11}.  
Here we illustrate the impact of the three-gluon correlation function on the asymmetry 
$A_N^\gamma\equiv \Delta\sigma^\gamma/\sigma^\gamma$ 
based on the same models used in the study of $p^\uparrow p\to DX$.

In \cite{KY11}, we have calculated $A_N^D$ at the RHIC energy with the two models for
$O(x)$ and $N(x)$:
(Model 1); $O(x)=K_GxG(x)$ and (Model 2); $O(x)=K'_G\sqrt{x}G(x)$ with the 
gluon density in the nucleon $G(x)$ assuming $O(x)=N(x)$. 
The constants $K_G$ and $K'_G$ were determined to be $K_G=0.004$ and $K'_G=0.001$
so that the calculated $A_N^D$ are consistent with the preliminary data of RHIC.  
\footnote{In \cite{KY11}, we employed an ansatz, $O(x,x)=K_GxG(x)$ and $O(x,x)=K'_G\sqrt{x}G(x)$
with the relation $O(x,x)=O(x,0)=N(x,x)=-N(x,0)$ and thus $K_G$ and $K'_G$ in this paper are
twice as large as those in \cite{KY11}.}
For the case of $D$-meson, change of the relative sign between $O(x)$ and $N(x)$
causes the interchange of $A_N^D$ between $D$ and $\bar{D}$ mesons.  
To calculate $A_N^\gamma$, we also use the above two models for the two cases:
(Case 1); $O(x)=N(x)$ and 
(Case 2); $O(x)=-N(x)$. 
We use the GJR08 \cite{GJR08} for the unpolarized distributions in the nucleon
and calculate $A_N^\gamma$ as a function of $x_F=2q_\parallel/\sqrt{S}$ at
the RHIC energy of $\sqrt{S}=200$ GeV and the
transverse momentum of the photon $q_T=2$ GeV, 
setting the scale of all the distribution functions at $\mu=q_T$.

\begin{figure}[h]
\begin{center}

\vspace{-0.5cm}
\scalebox{0.55}{\includegraphics{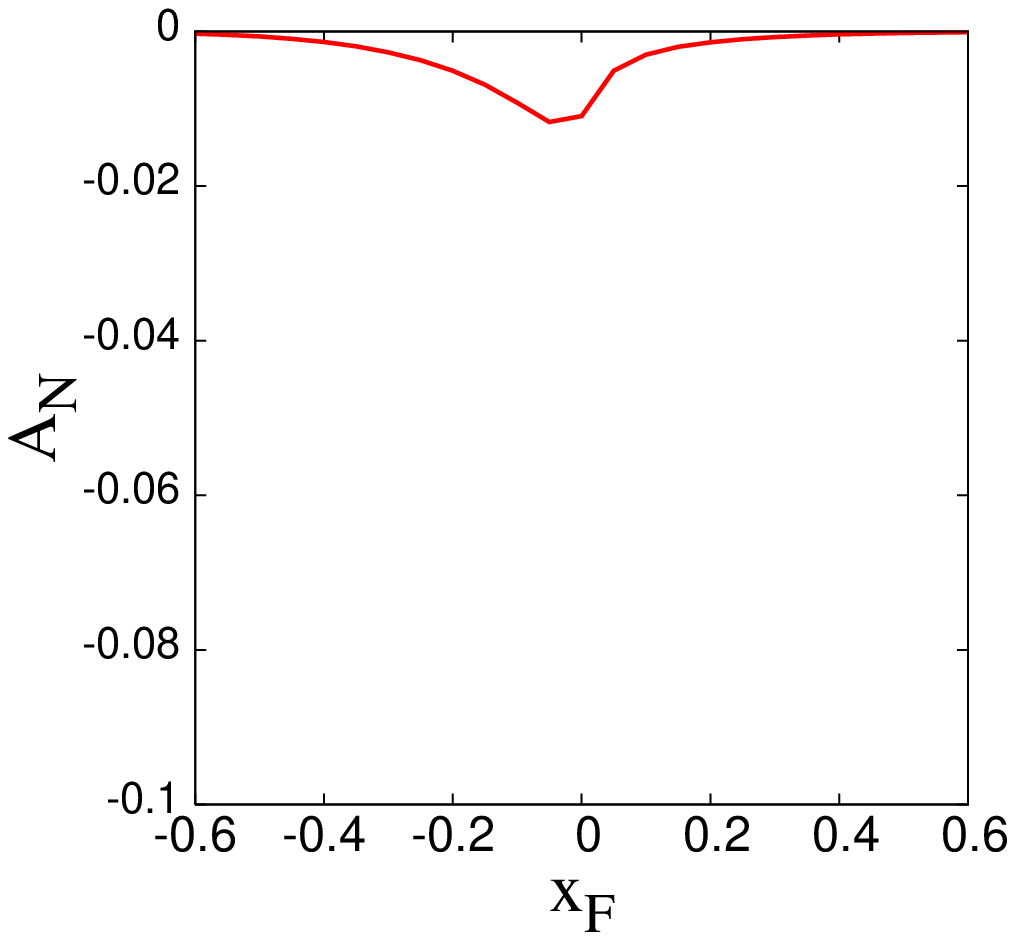}}
\scalebox{0.55}{\includegraphics{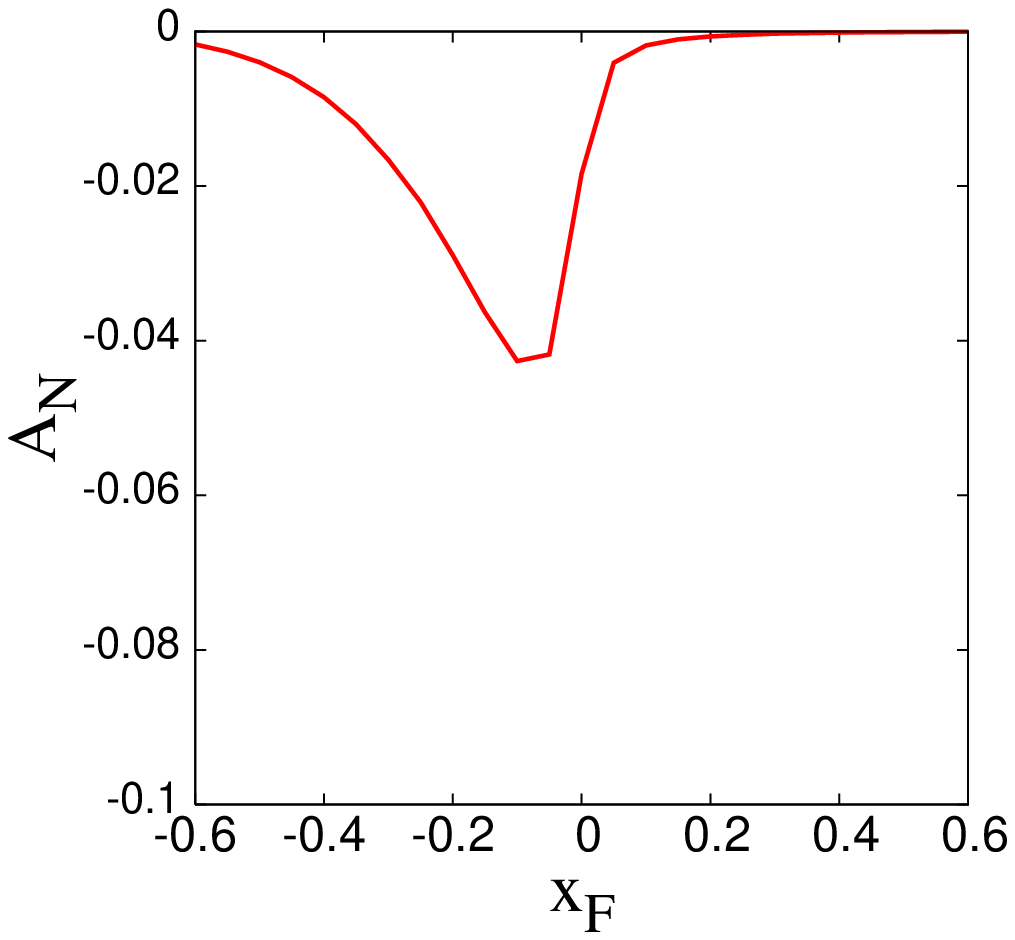}}

\hspace{0.7cm}(a)\hspace{6.3cm}(b)

\vspace{0.3cm}

\scalebox{0.55}{\includegraphics{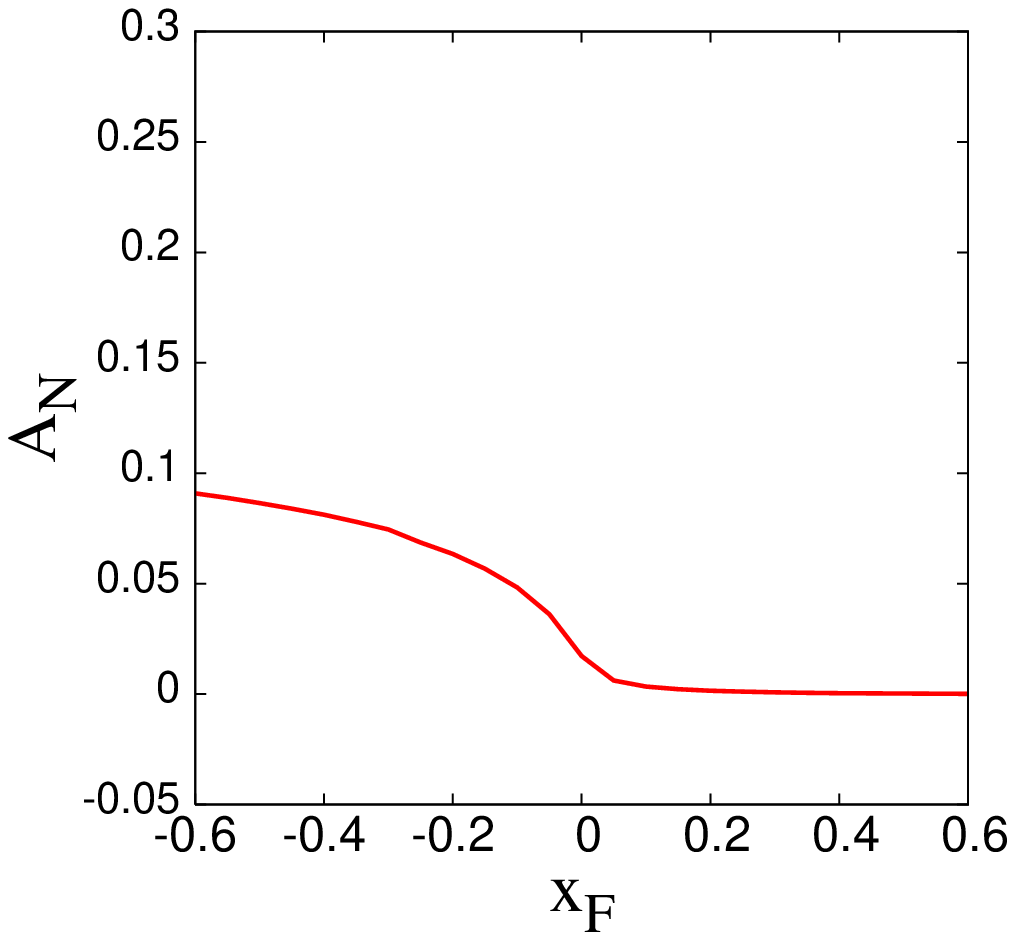}}
\scalebox{0.55}{\includegraphics{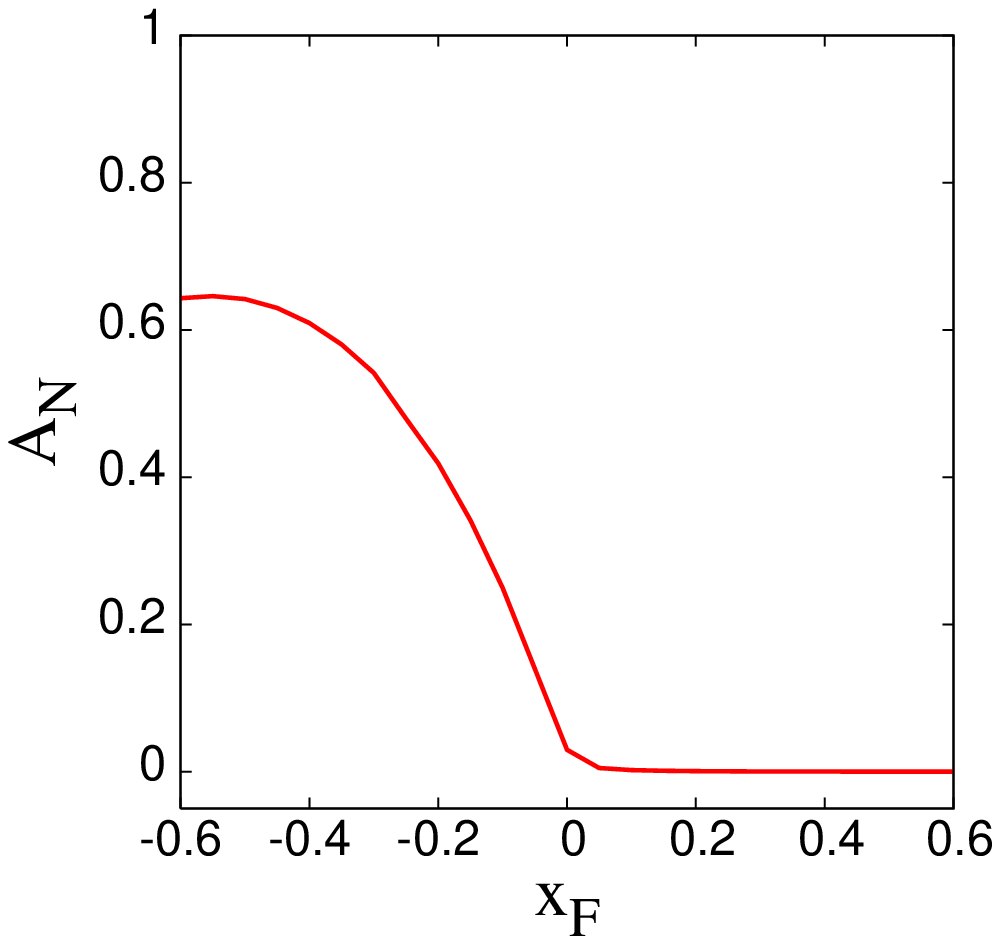}}

\hspace{0.7cm}(c)\hspace{6.3cm}(d)

\caption{$A_N^\gamma$ as a function of $x_F$ at the RHIC energy $\sqrt{S}=200$ GeV and $q_T=2$ GeV. 
(a) $A_N^\gamma$ for the case 1 with the model 1. (b) $A_N^\gamma$ for the case 1 with the model 2. 
(c) $A_N^\gamma$ for the case 2 with the model 1. (d) $A_N^\gamma$ for the case 2 with the model 2.}

\end{center}
\end{figure}

Fig. 3 shows the result for $A_N^\gamma$ for each case. One can see
$A_N^\gamma$ at $x_F> 0$ become almost zero regardless of the magnitude of
the three-gluon correlation functions, 
while $A_N^\gamma$
at $x_F<0$ depends strongly on the relative sign between $O(x)$ and $N(x)$ and also
the small-$x$ behavior of the
three-gluon correlation functions as in the case of $p^{\uparrow}p\to DX$.
Even though the derivatives of $O(x)$ and $N(x)$ contribute, 
$A_N^\gamma$ is tiny at $x_F>0$ due to the small partonic cross section
which occurs from the $u$-channel diagrams.  
At $x_F<0$, large-$x'$ region of the unpolarized quark distributions
and the small-$x$ region of the three-gluon correlation functions are relevant.  
For the above case 1, only antiquarks in the unpolarized nucleon are active
and thus leads to small $A_N^\gamma$ as shown in Figs. 3(a) and (b).  
On the other hand, 
for the case 2, quarks in the unpolarized nucleon are active and thus
lead to large $A_N^\gamma$ as shown in  Figs. 3(c) and (d).  
Therefore $A_N^\gamma$ at $x_F<0$ for the direct photon
production could provides us with an important information on the relative sign between $O(x)$
and $N(x)$.

In the present calculation of $A_N^\gamma$, we have included only the three-gluon correlation
functions for the spin-dependent cross section.  For the complete calculation,
one has to include the contribution from the quark-gluon correlation functions.
The quark-gluon correlation functions, however, gives rise to $A_N^\gamma$ only in the positive $x_F$
region\,\cite{kanazawa}
as in the case of $A_N$ for $p^\uparrow p\to \pi X$.  
Therefore, in $A_N^\gamma$, the role of three-gluon correlation functions and the quark-gluon correlation functions
are clearly separated and $A_N^\gamma$ at
$x_F<0$ truely plays an important role to
probe the three-gluon correlation functions.

\section{Summary}

In this paper, we have derived the twist-3 single-spin-dependent
cross section for the Drell-Yan lepton-pair production and the
direct-photon production in the $pp$ collision
induced by the three-gluon correlation functions
in the LO QCD.  The derivation was carried out by 
using the master formula for the three-gluon correlation functions,
which simplifies the calculation and makes connection between the
corresponding hard cross section and the $gq\to\gamma^{(*)}q$
hard scattering in the twist-2 level.  
The cross section for the Drell-Yan process
receives the contribution from the four functions $O(x,x)$, $O(x,0)$, $N(x,x)$
and $N(x,0)$ with different hard cross sections as in the
case of the SIDIS, $ep^\uparrow\to eDX$.  
For the direct-photon production,
the cross section is expressed in terms of the particular combinations
$O(x,x)+O(x,0)$ and $N(x,x)-N(x,0)$ with the common hard cross section which is
the same as the twist-2 unpolarized hard cross section in the $gq\to\gamma q$ 
scattering channel.  This simplification
is the same as those for the three-gluon contribution to $p^\uparrow p\to DX$
in the $m_c\to 0$ limit\,\cite{KY11} and the SGP contribution of the quark-gluon correlation 
function to $p^\uparrow p\to \pi X$\,\cite{KT072}.  
To illustrate the impact of three-gluon correlation functions,
we have performed a model calculation for SSA in the direct-photon
production $A_N^\gamma$, using the model determined from the SSA data
for $p^\uparrow p\to DX$ at RHIC.
It turned out that the effect of the three-gluon correlation functions 
is negligible in the forward direction of the polarized nucleon,
but can be substatial in the backward direction, depending on the small-$x$
behavior of the three-gluon correlation functions and the relative sign
of the two functions $O(x,x)+O(x,0)$ and $N(x,x)-N(x,0)$. 
Since the effect of the quark-gluon correlation on SSA is negligible
in the backward direction, the SSA in that region provides us with an important information
on the three-gluon correlation.

\vspace{1cm}

\noindent
{\bf Acknowledgement}

This work is supported by the Grand-in-Aid for Scientific Research
(No. 23540292 and No. 22.6032) from the Japan Society for the Promotion of Science.



\begin{thebibliography}{99}


\bibitem{review}
U. D'Alesio, F. Murgia, Prog. Part. Nucl. Phys. 61 (2008) 394;\\ 
V.~Barone, F.~Bradamante, A.~Martin, Prog. Part. Nucl. Phys. 65 (2010) 267.  

\bibitem{ET82} A.~V. Efremov and O.~V. Teryaev, Sov. J. Nucl. Phys. {\bf 36} (1982) 140
[Yad. Phiz. {\bf 36} (1982) 242]; Phys. Lett. {\bf B150} (1985) 383.

\bibitem{QS92} J. Qiu and G. Sterman, Nucl. Phys. {\bf B378} (1992) 52.

\bibitem{QS99}
J. Qiu and G. Sterman,
Phys. Rev. {\bf D59} (1998) 014004.

\bibitem{EKT07} H. Eguchi, Y. Koike and K. Tanaka, Nucl. Phys. {\bf B763} (2007) 198.

\bibitem{KK00} Y. Kanazawa and Y. Koike, Phys. Lett. {\bf B478} (2000) 121;
Phys. Lett. {\bf B490} (2000) 99; Phys. Rev. {\bf D64} 034019 (2001). 

\bibitem{EKT06} 
H. Eguchi, Y. Koike and K. Tanaka,
Nucl. Phys. {\bf B752} (2006) 1. 

\bibitem{JQVY06} X.~D.~Ji, J.~W.~Qiu, W.~Vogelsang and F.~Yuan,
Phys. Rev. Lett. {\bf 97} (2006) 082002; Phys. Rev. {\bf D73} (2006) 094017.

\bibitem{JQVY06DIS} X.~D.~Ji, J.~W.~Qiu, W.~Vogelsang and F.~Yuan,
Phys. Lett. {\bf B638} (2006) 178. 

\bibitem{KVY08}
  Y.~Koike, W.~Vogelsang and F.~Yuan,
  Phys.\ Lett.\  B {\bf 659} (2008) 878. 

\bibitem{Kouvaris}
 C.~Kouvaris, J.~W.~Qiu, W.~Vogelsang and F.~Yuan,
Phys.\ Rev.\  D {\bf 74} (2006) 114013. 

\bibitem{KT071}
  Y.~Koike and K.~Tanaka,
  Phys.\ Lett.\  B {\bf 646} (2007) 232
  [Erratum-ibid.\  B {\bf 668} (2008) 458]

\bibitem{KT072}
  Y.~Koike and K.~Tanaka,
  Phys.\ Rev.\  D {\bf 76} (2007) 011502
  
\bibitem{Tomita09}
  Y.~Koike and T.~Tomita, Phys. Lett. {\bf B675} (2009) 181. 

\bibitem{KK10}
  K.~Kanazawa and Y.~Koike,
  Phys.\ Rev.\  D {\bf 82}, 034009 (2010); Phys.\ Rev.\ D\ {\bf 83}, 114024  (2011)
  
\bibitem{Metz:2010xs}
  A.~Metz and J.~Zhou,
  Phys.\ Lett.\ B\ {\bf 700}, 11  (2011)

\bibitem{KK11} 
  K.~Kanazawa and Y.~Koike,
Phys.\ Lett.\ B\ {\bf 701}, 576  (2011)

\bibitem{MSZ11} 
  J.~P.~Ma, H.~Z.~Sang and S.~J.~Zhu,
arXiv:1111.3717 [hep-ph].

\bibitem{Ji92} X. Ji, Phys. Lett. {\bf B289}, 137 (1992).

\bibitem{KQ08}
  Z.~B.~Kang and J.~W.~Qiu,
  Phys.\ Rev.\  D {\bf 78} (2008) 034005.
  
\bibitem{KQVY08}  
  Z.~B.~Kang, J.~W.~Qiu, W.~Vogelsang and F.~Yuan,
  Phys. Rev. {\bf D78}, 114013 (2008).  

\bibitem{BKTY10}
  H.~Beppu, Y.~Koike, K.~Tanaka and S.~Yoshida,
  Phys.\ Rev.\  D {\bf 82}, 054005 (2010). 

\bibitem{KTY11}
  Y.~Koike, K.~Tanaka and S.~Yoshida,
  Phys.\ Rev.\ D\ {\bf 83}, 114014  (2011)

\bibitem{KY11} 
  Y.~Koike and S.~Yoshida,
Phys.\ Rev.\ D\ {\bf 84}, 014026  (2011).  

\bibitem{KY112} 
  Y.~Koike and S.~Yoshida, 
  in the proceedings of the XIX International Workshop on Deep-Inelastic Scattering and Related Subjests(DIS2011), 
  Virginia, USA, 
Apr.11-15, 2011, 
arXiv:1107.0512 [hep-ph].
  
\bibitem{KY113} 
  Y.~Koike and S.~Yoshida, in the proceedings of
the 8th Circum-Pan-Pacific Symposium 
on High Energy Spin Physics (PacSPIN2011), Cairns, Australia, June 20-24, 2011, 
arXiv:1110.6496 [hep-ph].

\bibitem{Liu}
  H.~Liu  [PHENIX Collaboration],
  AIP Conf.\ Proc.\  {\bf 1149} (2009) 439.
  
\bibitem{Jpsi} 
  A.~Adare {\it et al.} [PHENIX Collaboration],
Phys.\ Rev.\ D\ {\bf 82}, 112008  (2010)

\bibitem{BJLO01} A.~V. Belitsky, X.~D. Ji, W. Lu, J. Osborne, Phys.\ Rev.\ {\bf D63}, 094012 (2001). 

\bibitem{Braun09}
V.M. Braun, A.N. Manashov, B. Pirnay, Phys. Rev. {\bf D80}, 114002 (2009). 

\bibitem{GJR08} M. Gluck, P. Jimenez-Delgado, and E. Reya, 
	Eur. Phys. J. {\bf C53} 355 (2008). 
	
\bibitem{kanazawa}
K. Kanazawa, private communication.

\end{thebibliography}
\end{document}